\documentclass[final]{svjour3}
\usepackage{graphicx}
\usepackage{rotating}
\usepackage{amssymb}
\usepackage{mathptmx}
\usepackage[numbers]{natbib}
\usepackage{bibentry}
\journalname{Journal of Low Temperature Physics}

\bibpunct{}{}{,}{s}{}{,}

\begin{document}

\newcommand{\hdblarrow}{H\makebox[0.9ex][l]{$\downdownarrows$}-}
\title{Thermal simulations of temperature excursions on the Athena X-IFU detector wafer from impacts by cosmic rays}
\titlerunning{Thermal simulations of cosmic ray energy propagation in the X-IFU detector wafer}

\author{S. L. Stever$^{1, 2, 3}$ \and P. Peille$^{4}$  \and M. P. Bruijn$^{5}$ \and A. Roussafi$^{2}$, \and S. Lotti$^{6}$ \and C. Macculi$^{6}$ \and R. M. J. Janssen$^{2, 7}$ \and R. den Hartog$^{5}$}
\authorrunning{S. L. Stever et al.}

\institute{
\textbf{[1.]} Kavli IPMU (WPI), UTIAS, The University of Tokyo, Kashiwa, Chiba 277-8583, Japan\\
\textbf{[2.]} Institut d'Astrophysique Spatiale, \textit{INSU/CNRS}, Bt. 121, Universit\'e Paris-Sud,  Orsay, 91405, France\\
\textbf{[3.]} Laboratoire de l'Acc\'el\'erateur Lin\'eaire, \textit{IN2P3/CNRS}, Bt. 200,  Universit\'e Paris-Sud,  Orsay, 91405, France\\
\textbf{[4.]} Centre National d'\'Etudes Spatiales (CNES), 18 Avenue Edouard Belin, 31400 Toulouse, France\\
\textbf{[5.]} Netherlands Institute for Space Research (SRON), Sorbonnelaan 2, 3584 CA Utrecht, Netherlands\\
\textbf{[6.]} INAF/IAPS Roma, Via del Fosso del Cavaliere 100, 00133 Roma, Italy\\ 
\textbf{[7.]} NASA Jet Propulsion Laboratory, 4800 Oak Grove Dr., Pasadena CA 91109, United States of America\\ 
\email{samantha.stever@ipmu.jp}}

\maketitle

\begin{abstract}

We present the design and implementation of a thermal model, developed in COMSOL, aiming to probe the wafer-scale thermal response arising from realistic rates and energies of cosmic rays at L2 impacting the detector wafer of Athena X-IFU. The wafer thermal model is a four-layer 2D model, where 2 layers represent the constituent materials (Si bulk and Si$_{3}$N$_{4}$ membrane), and 2 layers represent the Au metallization layer's phonon and electron temperatures. We base the simulation geometry on the current specifications for the X-IFU detector wafer, and simulate cosmic ray impacts using a simple power injection into the Si bulk. We measure the temperature at the point of the instrument's most central TES detector. By probing the response of the system and pulse characteristics as a function of the thermal input energy and location, we reconstruct cosmic ray pulses in Python. By utilizing this code, along with the results of the GEANT4 simulations produced for X-IFU, we produce realistic time-ordered data (TOD) of the temperature seen by the central TES, which we use to simulate the degradation of the energy resolution of the instrument in space-like conditions on this wafer. We find a degradation to the energy resolution of 7 keV X-rays of $\approx$0.04 eV. By modifying wafer parameters and comparing the simulated TOD, this study is a valuable tool for probing design changes on the thermal background seen by the detectors.

\keywords{cosmic ray, particle interactions, X-ray, systematic effects}

\end{abstract}
\vspace{-2em}
\section{Introduction}
The X-Ray Integral Field Unit (X-IFU), aboard the future Advanced Telescope for High Energy Astrophysics (Athena), was selected in 2014 by the European Space Agency to answer the call of the Hot and Energetic Universe scientific theme [1]. X-IFU will measure high-resolution X-ray spectra between 0.2 to 12 keV, with a spectral resolution of 2.5 eV for $E$ $<$ 7 keV and $E/dE$ = 2800 for $E >$ 7 keV. Its science targets include physical characterization of galactic winds, AGN, plasma, spectroscopic line detection from galactic filaments, and mapping and characterization of hot plasma [2]. These targets drive a variety of instrument goals, including the spectral resolution.\\

Given X-IFU's ambitious scientific requirements, precise understanding and control over systematic effects is necessary. This manuscript will focus on the effect of cosmic ray impacts, particularly into the X-IFU detector wafer, on the energy resolution of the instrument. Cosmic Rays (CRs), particularly protons, will interact with the spacecraft situated at the second Earth-Sun Lagrange Point (L2). These protons will penetrate the spacecraft and create secondary particles, depositing energy wherever they go: the detectors, the wafer, or other mechanical structures. This effect becomes significant at this level of instrument sensitivity, as was the case with the Planck Space mission [3]. This manuscript focusses specifically on the frequent impacts by CRs into the detector wafer, which create a baseline thermal fluctuation seen by the detectors, and whether the level of these fluctuations will affect the energy resolution of the instrument. The other potential sources of CR contamination are the topic of other studies.\\

In Sec.~\ref{sec:Model}, we will discuss a thermal model produced in COMSOL of the X-IFU detector wafer. In Sec.~\ref{sec:PyModel}, we will show how to reproduce the pulses produced by the thermal model for the central-most point of the wafer, or the $\Delta T(t)$ pulse seen by the most central detector, using a scaling model. This model takes input of energy depositions on the X-IFU wafer as simulated in GEANT4 using an X-IFU mass model under L2 radiation conditions. Using these results, we produce $\Delta T(t)$ time-ordered data (TOD), which are discussed in Sec.~\ref{sec:Results}, alongside the effect of this TOD on a 7 keV X-rays in the \textit{xifusim} instrument simulator.\\

We note that the goal of this simulation is to determine the baseline level of heating of the wafer from many concurrent CRs, in addition to the effect of this heating on the energy resolution. In this, we consider only diffusive thermal transport due to a lack of athermal phonon sim- ulation mechanisms within COMSOL. A simulation also including energy deposition into a TES would require both thermal and athermal energy transport and will be the topic of other studies.\\
\vspace{-2em}
\section{COMSOL Thermal Model}
\label{sec:Model}

\begin{figure}[htbp]
\centering
\includegraphics[width=0.99\linewidth, keepaspectratio]{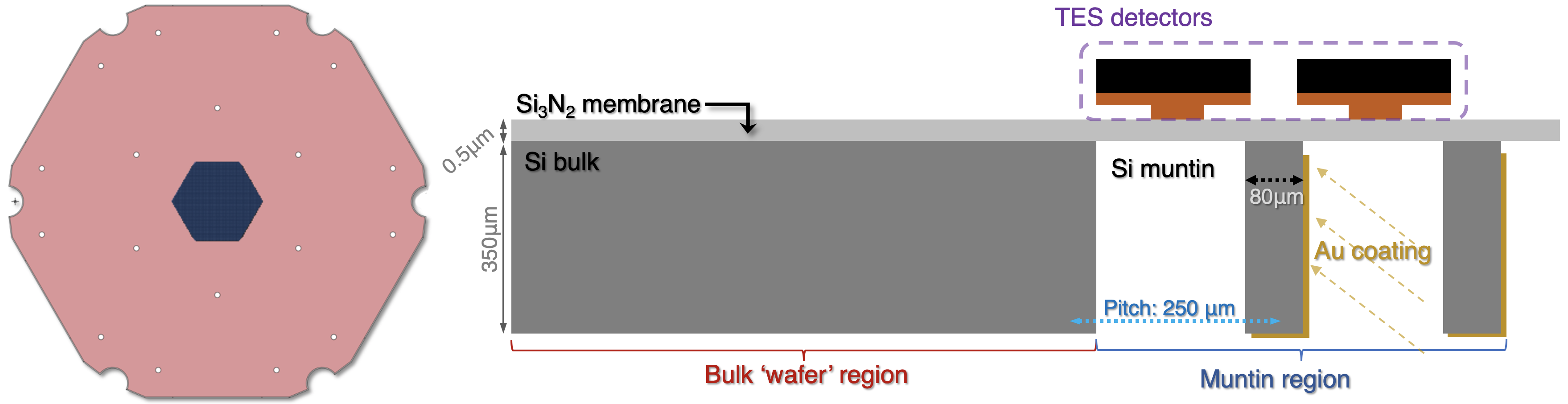}
\caption{\textit{Left:} CAD model of the detector wafer, with the `wafer region' (red) and the `muntin region' (blue). \textit{Right:} Side view of the wafer geometry, showing the wafer and muntin regions. The TES (encircled in dashed purple) are not included in the thermal model. (Colour figure online.)}
\label{img:fig1_new}
\end{figure}

The wafer model has been produced in COMSOL Multiphysics version 5.3a with the Heat Transfer module, and is a 2D four-layer finite-element model with virtual $z$-axis thicknesses, based on the current design for the X-IFU detector wafer. The X-IFU wafer has two distinct regions as shown in Fig.~\ref{img:fig1_new}: the `wafer' region acts as a thick support layer, which also houses the wirebonds, and the `muntin' structure contains a fine grid of bulk-removed Si, where the detectors sit upon a thin membrane of Si$_{3}$N$_{4}$. The Au metallization below the grid is produced by two spray patterns at two angles, and is represented by two model layers - one for the phonon temperature and one for the electron temperature. The model is based upon a prior model of the SPICA SAFARI [4] wafer developed at SRON, which we have adapted for X-IFU. \\

\begin{figure}[htbp]
\centering
\includegraphics[width=0.6\linewidth, keepaspectratio]{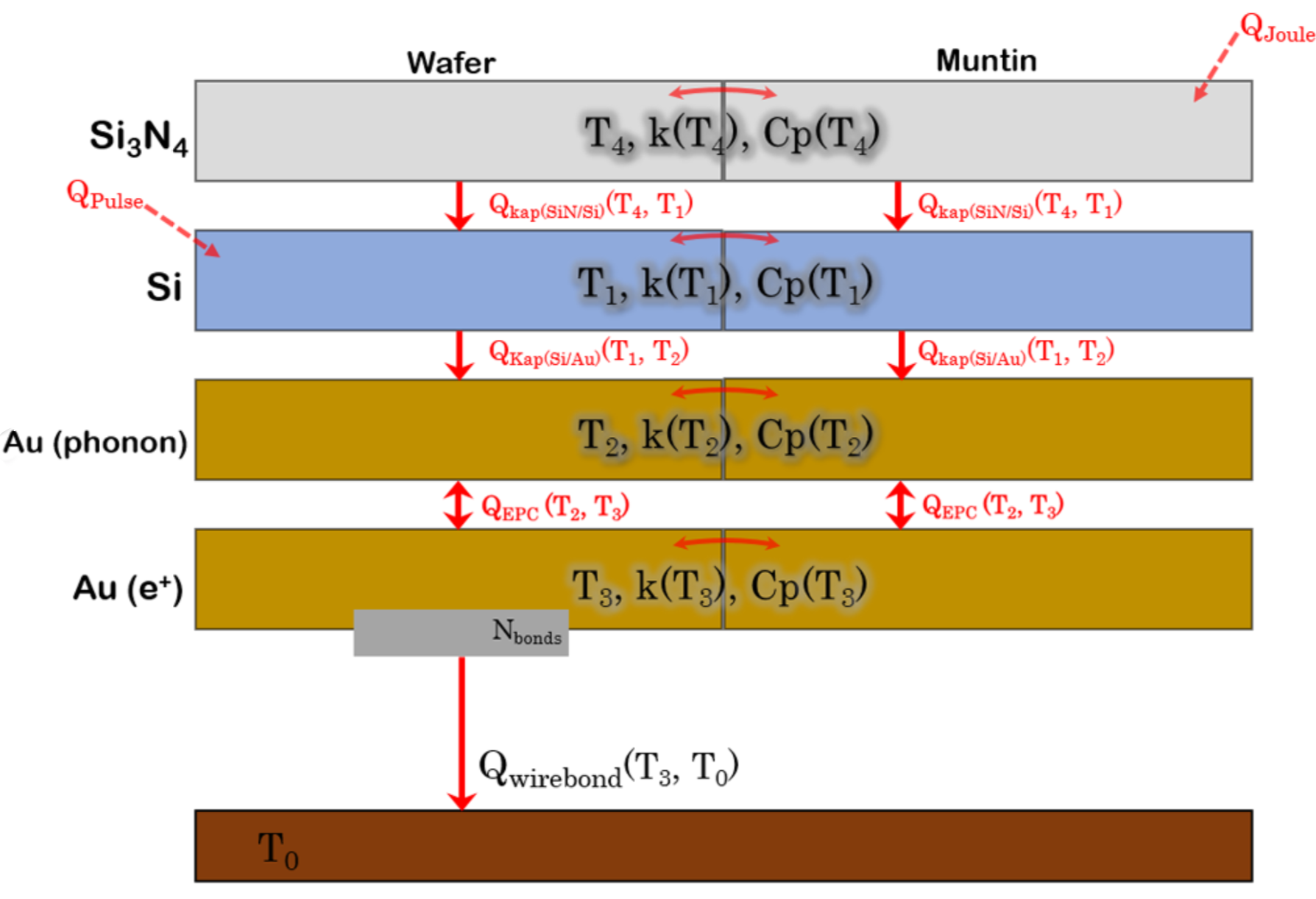}
\caption{Block diagram of the thermal model, with each block representing a different layer in the COMSOL model.  (Colour figure online.)}
\label{img:fig2_new}
\end{figure}

We show a block diagram of the model in Fig.~\ref{img:fig2_new}. Each block represents a layer of material, each with its own thermal properties (e.g. heat capacity, thermal conductance) determined either from measurements or from the literature. The heat propagation between the material layers, arising from interfacial thermal resistance, is Kapitza-like and scales as $T^{4}$. The electron-phonon coupling between the Au phonon and electron layers scales as $T^{5}$. Although this simulation accounts only for the wafer and contains no TES detectors (which will be discussed in the next section), we account for the Joule heating induced by the TES readout by applying a uniform power $\times$ $N_{\textrm{det}}$ across the `muntin' structure of the geometry. A CR is then approximated as a fast exponential pulse with a rise time of the order of $10^{-9}$ s and a decay time to the order of $10^{-7}$ s (a Dirac delta form would be the best representation, but cannot be used in COMSOL). Thermal coupling to $T_{0}$ = 50 mK is done using a dedicated `wirebond region' geometry. A fine mesh is applied through the geometry, and a steady-state solution is produced across the wafer. The CR energy is injected into a chosen location, and the software solves for its propagation as a function of time.\\

In the Si, \textrm{Si$_{3}$N$_{4}$}, and Au phonon layers, the wirebonding region is uniformly integrated into the `wafer' region in the COMSOL geometry. In the Au electron layer, the wirebonding region uses the same thermal properties as the surrounding Au, attenuated by a filling factor calculated assuming that $(i)$ $N_{\textrm{bonds}}$ wirebonds are stacked next to each other around the perimeter and $(ii)$ that their $G$ does not have a gradient along its length. The number of wirebonds is undecided, and this simulation assumes $N_{\textrm{bonds}}$ = 100. Their thermal conductance and geometrical parameters are set from measurements of a single gold wirebond performed at NASA Goddard [5].\\

\begin{figure}[htbp]
\centering
\includegraphics[width=0.7\linewidth, keepaspectratio]{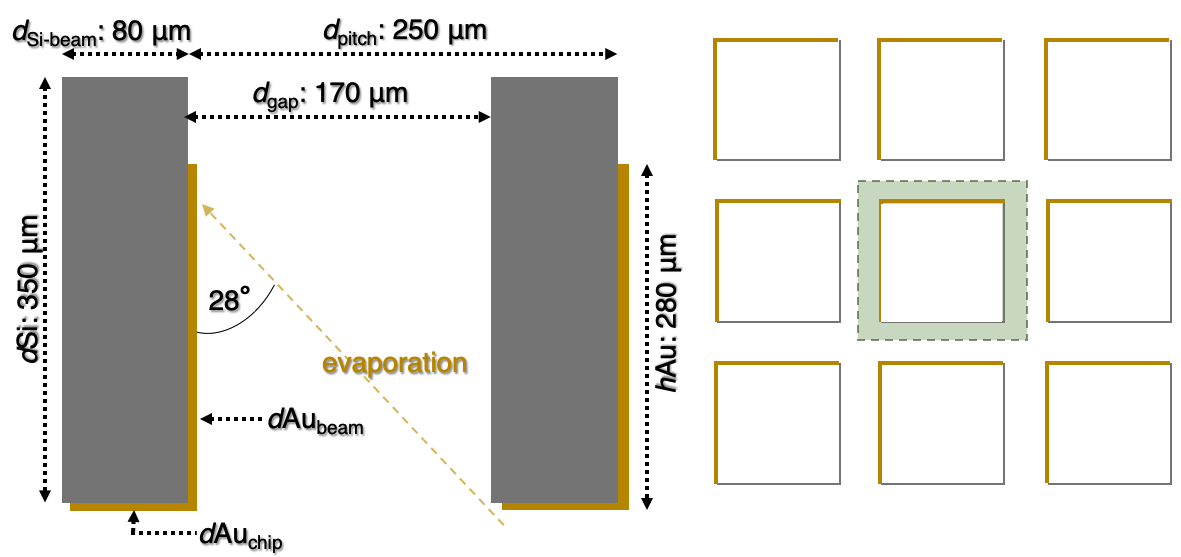}
\caption{\textit{Left}: Two muntin beams and their geometrical parameters (Si in gray and Au in gold). \textit{Right}: Top-down view of 9 etched-out regions between the Si beams. The green shows one `unit cell' used in the Muntin approximation. (Colour figure online.)}
\label{img:muntin}
\end{figure}

Finally, the model makes use of a so-called `muntin approximation'; in the COMSOL simulation, we do not account for the full complexity of the muntin structure, which is computationally prohibitive. As we show in Fig.~\ref{img:muntin} (left), the evaporation of the Au is perpendicular to the beams in $x$ and then perpendicular to the beams in $y$, with an evaporated thickness of $d_{\textrm{evap}}$ = 1.7 $\mu$m. The metal is deposited at an angle of 28$^{\circ}$ relative to $y~=~0$. The thickness of the deposited gold on the bottom of the beam is $d$Au$_{\textrm{chip}}$ and the thickness on the sides is $d\textrm{Au}_{\textrm{beam}}$. These thicknesses are defined by:

\begin{equation}
d\textrm{Au}_{\textrm{chip}} = 2 \cdot d_{\textrm{evap}} \cdot \cos(\theta)
\label{dCuchip}
\end{equation}

\noindent where $\theta = \tan^{-1}$($d_{\textrm{gap}}$/$h_{\textrm{Au}}$), $d_{\textrm{gap}}$ is the width of the gap between two Si beams, and $h_{\textrm{Au}}$ is the height of the deposited Au. The thickness of the Au on the sides of the beams, $d$Au$_{\textrm{beam}}$, is simply $d_{\textrm{evap}}$ $\cdot \sin(\theta)$. To account for this in the thickness (compensating in the electron-phonon coupling, thermal conductance, and heat capacity) we define an effective thickness of the Au on the sides of the beams which transfers the thickness of the metal evaporated on the beams to their bottom:

\begin{equation}
d\textrm{Au}_{\textrm{eff}} = \frac{d\textrm{Au}_{\textrm{chip}} + 2 \cdot d\textrm{Au}_{\textrm{beam}} \cdot h_{\textrm{Au}} \cdot d_{\textrm{gap}}}{d_\textrm{Si-beam}(d_{\textrm{gap}} + d_{\textrm{pitch}})}
\label{dCu_eff}
\end{equation}

\noindent where $d_{\textrm{Si-beam}}$ is the width of the Si beams. The interfacial coupling relies on effective area rather than thickness, therefore we split the Si muntins into `cells' where $A_{\textrm{cell}}$ = $d_{\textrm{Si-beam}} \cdot (d_{\textrm{gap}}+d_{\textrm{pitch}})$, and the area of the Au coating where $A_{\textrm{coated}}$ = $A_{\textrm{cell}}$ + $2 \cdot h_{\textrm{Au}}\cdot d_{\textrm{gap}}$. The ratio of the Au-coated area with the area of each `cell' is the effective Kapitza surface (as shown in the right of Fig.~\ref{img:muntin}):

\begin{equation}
A\textrm{Au}_{\textrm{eff}} = \frac{(A_{\textrm{cell}} + 2 \cdot d_{\textrm{gap}} \cdot h_{\textrm{Cu}})}{A_{\textrm{cell}}}
\label{ACu_eff}
\end{equation}

In this way we can simulate using a simplified geometry, whilst accounting for the effect of the air gaps on the thermal parameters. \\
\vspace{-2em}
\subsection{Thermal Model Output}
\begin{figure}[htbp]
\centering
\includegraphics[width=0.7\linewidth, keepaspectratio]{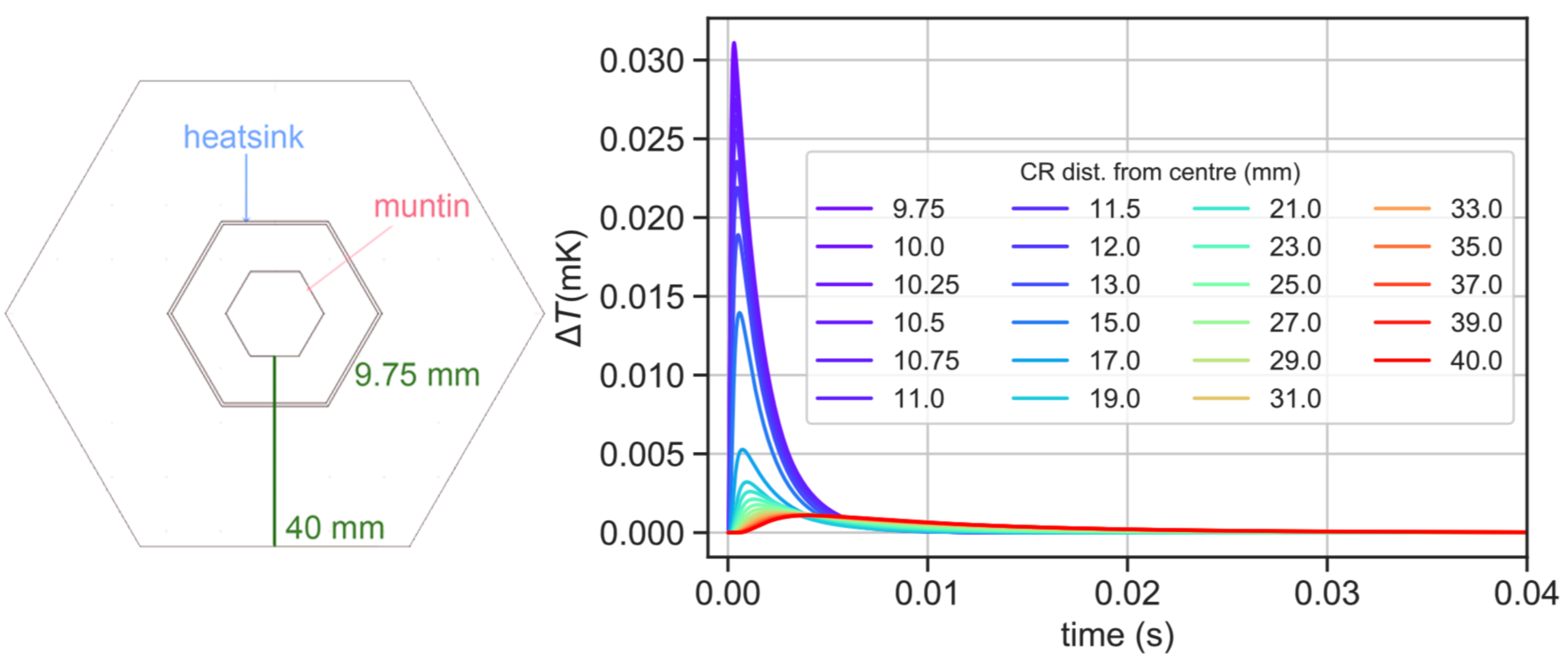}
\caption{Results of the thermal model, for a 500 keV heat pulse at locations varying between 9.75 to 40 mm from the central-most pixel (0,0). \textit{Left:} Diagram of the wafer, showing the location of the pixel and muntin regions, heatsink, and representation of varying CR injection distances (green line). (Colour figure online.)}
\label{img:Tfull_500keV_sink}
\end{figure}

The COMSOL model produces $\Delta T(t)$ curves for one set of initial parameters (CR energy and location) for the temperature of the top Si$_{3}$N$_{4}$ layer at the centre of the muntins, the temperature which the detector considers to be the thermal bath. We can then produce pulse libraries for temperature excursions by sweeping over the CR location parameter. We show such results in Fig.~\ref{img:Tfull_500keV_sink}, where a 500 keV simulated cosmic ray's starting position is varied along the $y$-axis, producing variable $\Delta T$ excursions on the central pixel. The image also shows the layout of the simulated wafer in $xy$ orientation. We find that CRs injected the closest to the centre of the wafer have the highest amplitudes, but also the shortest time constants (ms-scale), which is likely due to the influence of the nearby wirebonds. By contrast, the pulses at the periphery of the wafer have much lower amplitudes, but have time constants in the 10 to 20 ms range.\\

\begin{figure}[htbp]
\centering
\includegraphics[width=0.9\linewidth, keepaspectratio]{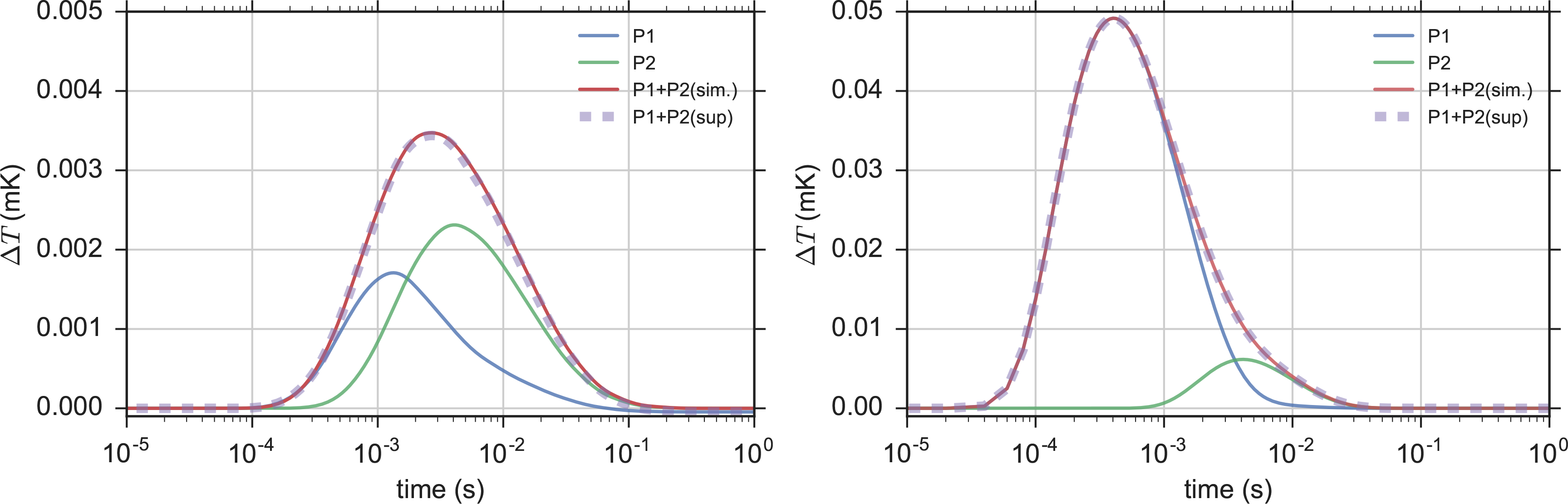}
\caption{Two superposition tests and their results. Pulses 1 (blue) and 2 (green) standalone, compared with the simulated run containing both pulses (red), and the superposition of the separately simulated pulses (purple dashes). (Colour figure online.)}
\label{img:superposition}
\end{figure}

Producing 83 second $\Delta T(t)$ TOD of space-like energy depositions in COMSOL is computationally limiting; 20 minutes per simulation for 119$\times$10$^{5}$ energy depositions would take $\approx$450 years to simulate one-by-one in COMSOL. However, we can utilize the scalability properties of the COMSOL output in order to quickly and reliably reproduce pulses and TOD. For this, it is necessary that $(i)$ artificial superposition of multiple pulses matches one simulation containing all of these pulses, and that $(ii)$ the pulses are scalable in relation to energy and distance. For this purpose, we have tested the COMSOL output twice by simulating 2 pulses separately; the first (P1) at a distance of -10 mm in $x$ and 20 mm in $y$ with an energy of 700 keV, and a second (P2)  at a distance of -12 mm in $x$ and 33 mm in $y$ with an energy of 2300 keV. This separate simulation is compared with a simulation containing both pulses, as well as the sum of both separate pulses in Fig.~\ref{img:superposition} (left). We repeat this test on two new pulses, where P1 has an $x$ location of 0 mm, a $y$ location of 11 mm, and an energy of 1000 keV. In the second test, P2 has a much higher energy of 3000 keV, but is farther away with an $x$ location of 35 mm and a $y$ location of -20 mm. We see very little difference between the two cases (with the largest residual constituting $\approx$0.35\% of the signal magnitude), showing we can rely upon the superposition between two separately simulated pulses, even if they are in very different locations.\\
\vspace{-2em}
\section{Production of cosmic ray TOD}
\label{sec:PyModel}

\begin{figure}[htbp]
\centering
\includegraphics[width=0.8\linewidth, keepaspectratio]{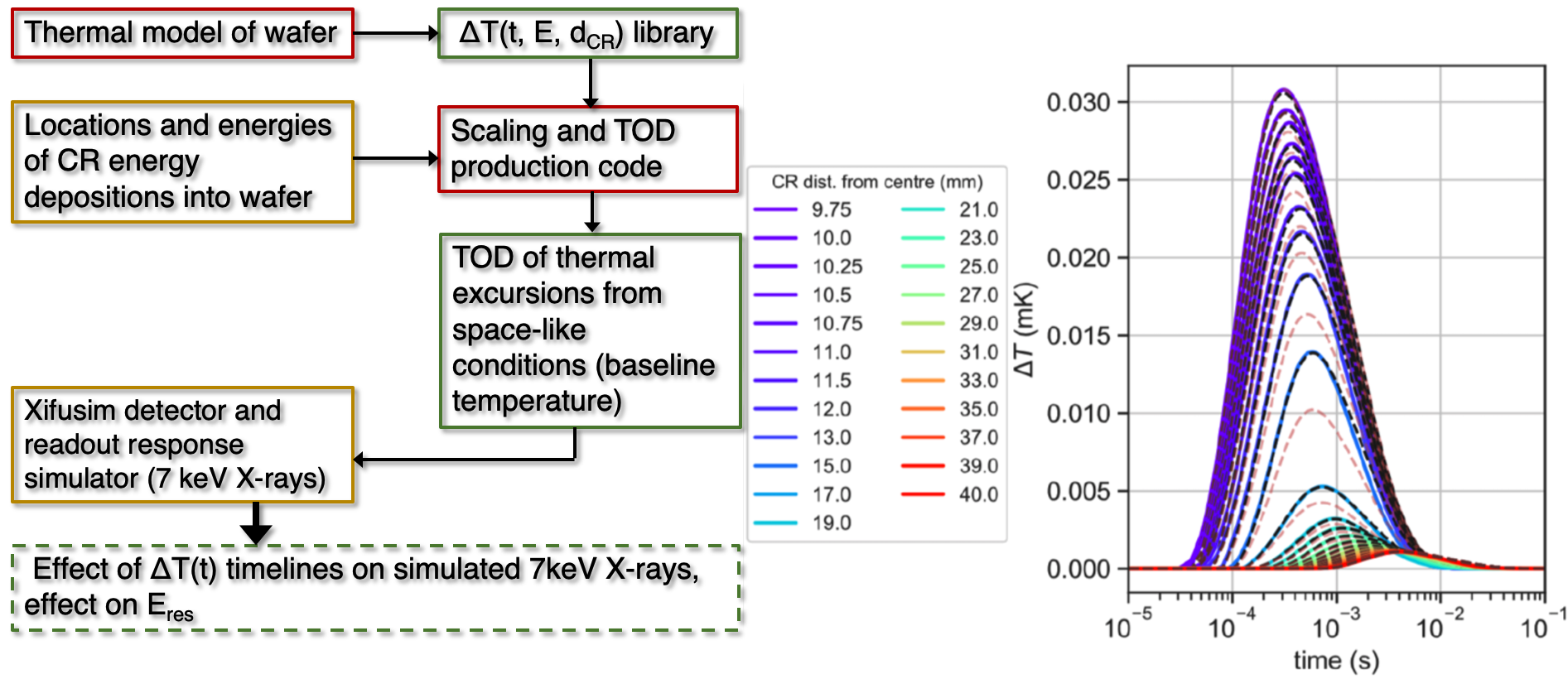}
\caption{\textit{Left}: Simulation workflow, showing the evolution from thermal model to simulated TOD and energy resolution impact estimations. \textit{Right, solid lines}: 500 keV simulated pulses; \textit{Right, black dashed lines}: 50 MeV pulses, downscaled to 500 keV; \textit{Right, dashed red lines}: Half-distance scaled and energy-scaled 50 MeV pulses. (Colour figure online)}
\label{img:pulse-scaling}
\end{figure}

The parameter sweeps in distance, with a constant CR energy, are saved as a pulse library. We produce 3 pulse libraries in total; cosmic ray energy levels of 500 keV, 5 MeV, and 50 MeV, which will be used in the Python modeling to produce simulated TOD. We use the pulse shape, which is determined by the distance, and then we scale it depending on the amplitude-energy relationships. The relationship between the thermal model pulse libraries, TOD production, and the final result is shown in the simulation workflow in Fig.~\ref{img:pulse-scaling} (left). \\

Next, the scaling code takes input from the results of a model produced in GEANT4 v.10.2 by INAF, of the X-IFU FPA and cryostat mass models with the forecasted CR environment at L2 (Pamela 2009 spectra) [6]. The GEANT4 data is based on the May 2017 mass model of X-IFU, and this analysis must be repeated when the next release becomes available. The GEANT4 simulation assumes an isotropic flux of primary protons that cross the mass of X-IFU and also produce showers of secondary particles (usually electrons). Both species hit the wafer, and GEANT4 is able to produce the spectra of the deposited energies. The outputs of the GEANT4 data are $x$, $y$, and $z$ locations of energy depositions from secondary particle showers into the X-IFU wafer over 83 s, along with the incident energy and deposited energy into the wafer. The average arrival energy of a secondary particle is $\approx$141 MeV, of which 0.58 keV is deposited into the wafer. Energies smaller than 200 keV, when simulated in COMSOL, led to precision errors due to the small size of the perturbation. We therefore expect that most energy depositions below the average value will create thermal excursions well below the noise, but we must downscale higher-energy pulses to simulate them. Each particle can have many secondary energy depositions (e.g. shallowly grazing minimally-ionising proton hits which leave small energy depositions along a track in the wafer, or widely-spread CR showers).\\

The primary loop iterates over each energy deposition, choosing the pulse closest to its energy and distance from the library. This pulse is normalised, and then rescaled as a function of the energy-amplitude relationships. It is further scaled in distance, if the distance of the library pulse is more than 25\% larger than the absolute value of its nearest-distance neighbors. The results of these scaling methods are shown in Fig.~\ref{img:pulse-scaling} (right), where we show 500 keV simulated pulses alongside downscaled 50 MeV pulses and their half-distance scaled counterparts. For every energy deposition, this process is repeated as we progress in time, where $dt$ of each primary energy deposition is drawn from the exponential distribution of the time difference between two events following Poisson statistics [7] using the calculated rate of 183 hits s$^{-1}$. The individual timestreams are added, and the final result is 83 seconds of $\Delta T(t)$ TOD at the Si$_{3}$N$_{4}$ membrane upon which the TES detector sits.\\

A simpler method to produce TOD is to use the `pre-analyzed' data, where all secondary energy depositions arising from one primary particle are added together and assigned to their average $x$ and $y$ location, as performed in Peille et al.[8], which we will refer to as the `primary deposition approximation'. This method decreases the number of total events from 1.13$\times$10$^{7}$ to 1.52$\times$10$^{3}$, but increases the average deposited energy from 0.58 to 452.17 keV. The computational time is decreased significantly (and it is possible to achieve 83 seconds of simulated data in $<$20 minutes). The drawback is that this is the least accurate treatment of the GEANT4 data, and may over-estimate the large-scale amplitudes in the TOD whilst underestimating the low-level thermal noise. In the next section, we will compare both methods using the \textit{xifusim} instrument simulator [9], which simulates X-ray data passing though the detector and the readout chain. Using the $\Delta T(t)$ TOD as the operating temperature for the TES, \textit{xifusim} simulates 500 7 keV X-rays, allowing us to determine the effect of these thermal excursions on the energy resolution of X-IFU.
\vspace{-1em}
\section{Results}
\label{sec:Results}
\begin{figure}[htbp]
\centering
\includegraphics[width=0.7\linewidth, keepaspectratio]{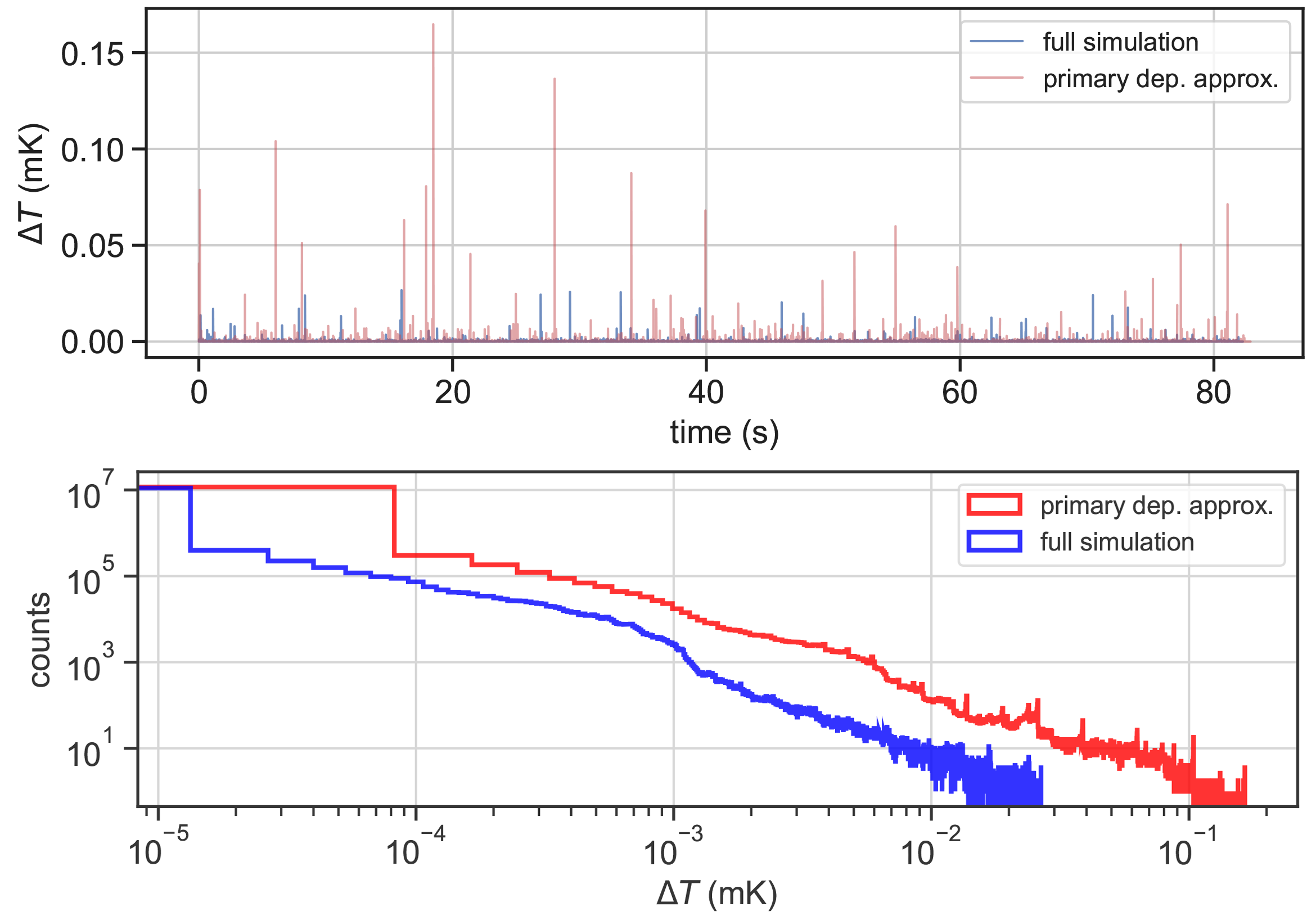}
\caption{\textit{Top}: The $\Delta T(t)$ TOD for the most central X-IFU pixel produced by processing the full simulation (blue) and the primary deposition approximation (red). \textit{Bottom}: Histograms of the above TOD.  (Colour figure online)}
\label{img:primaries_vs_secondaries}
\end{figure}

We show the results of both treatments in Fig.~\ref{img:primaries_vs_secondaries}. In the 83 second simulated TOD, we see that the primary deposition approximation creates higher-amplitude temperature excursions. This is seen in the RMS temperature of each TOD; we find that the $\Delta T_{\textrm{RMS}}$ of the primary deposition approximation data to be $1.64\times10^{-3}$ mK, compared with the fully-processed TOD which yield a $\Delta T_{\textrm{RMS}}$ of $3.28\times10^{-4}$ mK. We find, by comparing the histograms in Fig.~\ref{img:primaries_vs_secondaries} (bottom) that even the low-level thermal excursions are increased in the primary deposition approximation TOD; this is most likely due to the majority of the pulses arriving further away from the central pixel, and the longer time constants associated with them.\\

\begin{figure}[htbp]
\centering
\includegraphics[width=0.7\linewidth, keepaspectratio]{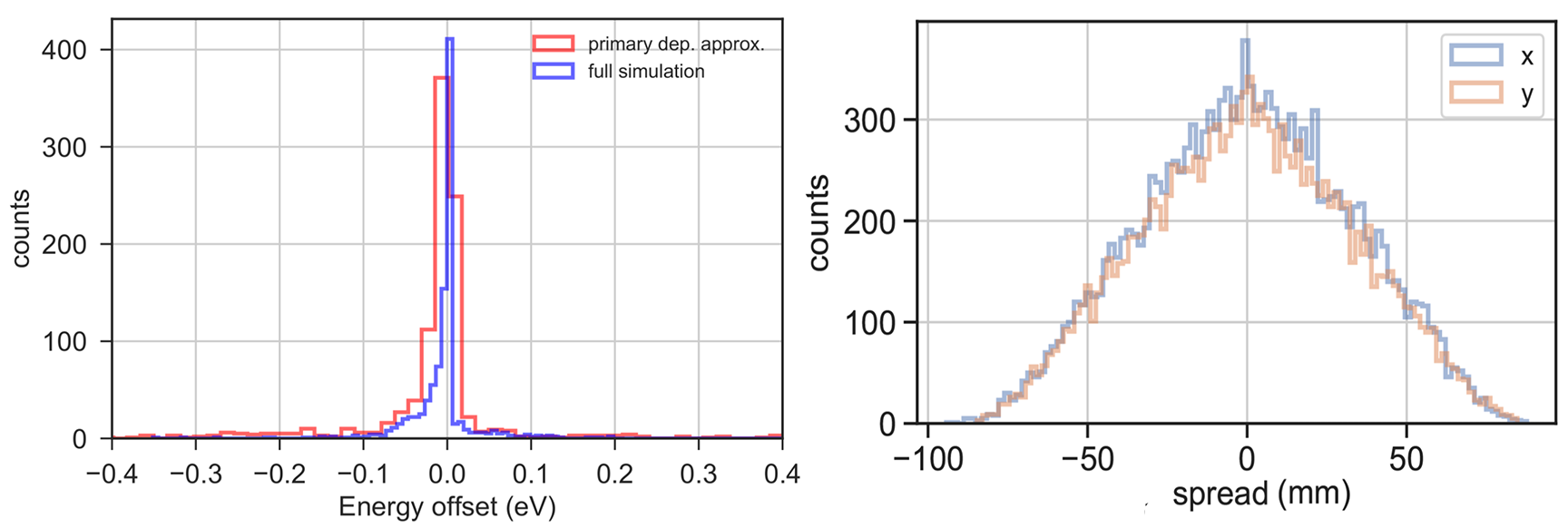}
\caption{\textit{Left:} Histogram of energy offsets of simulated 7 keV X-rays from the thermal excursions of the primary deposition approximation (red) and the full simulation (blue). \textit{Right:} The spread of secondary particles in each primary event (full simulation). (Colour figure online)}
\label{img:energyoffsets}
\end{figure}

By taking the simulated TOD as the TES operating temperature as a function of time, we can test the response of the TES to the wafer temperature fluctuation in \textit{xifusim}. This is done by simulating 500 7 keV X-rays from within \textit{xifusim} which solves the TES thermal equations, continuously taking the CR TOD as $T_{0}$, and calculating their reconstructed energies. We show the results in the histogram in Fig.~\ref{img:energyoffsets} (left). We find that for the primary deposition approximation (red), the FWHM of the energy offset is 0.06 eV. For the full (more accurate) simulation, the FWHM is 0.04 eV. Comparing the average reconstructed energy offsets with that of the budgeted degradation to the energy resolution of X-IFU, we find that both are below the budgeted level of 0.2 eV. Our results are in agreement with a parallel study by Peille et al. [7]. By checking the spread of the secondary particles in Fig.~\ref{img:energyoffsets} (right), we can conclude that the primary deposition approximation (which neglects the effect of showers and the spread of secondary particles) does not significantly impact the validity of the simulation. In this simulation, the effect of CR showers does not appear to be significantly different to that of more direct energy deposition within the regime of diffusive thermal transport.\\

\section{Discussion}
\label{sec:Discussion}
We have demonstrated a method of analyzing the effect of CRs arriving on the X-IFU detector wafer, and the effect of these depositions on the energy resolution of the instrument. We note that there are many means by which CRs can interfere with the energy resolution of the instrument: by directly impacting the detectors, by heating elements of the focal plane, in addition to a fluctuating baseline temperature in the wafer. The latter is just one potential effect. Although we have found that the overall extent of this effect is low compared with the budgeted level of degradation of the energy resolution, we have not yet considered hits in the muntin structure, nor in the detectors themselves, although this effect has been studied in Peille et al. [8]. Furthermore, a separate experiment and simulations will need to account for the propagation of ballistic phonons in the wafer, which will create an additional background, and is under study by INAF and the author.\\

Overall, the primary deposition approximation and the full simulation produce similar results, with the primary deposition approximation favoring larger-amplitude pulses as well as a stronger low-temperature component, which is likely due to the long time constants of the more frequent far-away pulses. One infers from this that the high-amplitude components of the temperature excursions have a larger effect on the energy resolution than the low-level thermal noise induced by the `secondary particle rain'. We see that the full data includes the effect of cosmic ray showers, which is an effect neglected in the primary deposition approximation, and does not appear to significantly affect the results. Therefore, we posit that the primary deposition approximation is a reasonable assumption which allows for the saving of computational time, and gives a `worst-case' scenario and therefore a conservative estimate of CR contamination. However, these simulations only demonstrate the method for the central-most detector region; one expects shorter time constants closer to the injection site of the CR, and longer time constants further from it. Future studies on edge pixels are a topic of upcoming simulations. Finally, as with all simulation output, this work must be verified experimentally, but can be used for probing instrument design changes.\\

\vspace{-3em}
\section{Conclusions}
We have simulated the effect of the cosmic ray environment at L2 on the detector wafer of the X-IFU instrument aboard the Athena telescope, using a Python model adapted from COMSOL, to produce $\Delta T(t)$ curves from 83 seconds of simulated data from GEANT4. We find, using two different methods of simulation, that the noise induced by cosmic rays on the central detector of X-IFU results in degradations of the energy resolution of 7 keV X-rays of 0.04 - 0.06 eV, which is below the total budgeted energy resolution degradation budget of 0.2 eV. This specific CR noise source is likely to act in tandem with other noise sources, e.g. direct detector hits, heavy ions, or FPA heating. However, the specific instance of wafer heating by cosmic rays in a worst-case L2-like environment is unlikely to be a significant threat to the energy resolution, although these results must be validated by a proton-beam experiment.\\

\begin{acknowledgements}
The corresponding author wishes to acknowledge CNES for their doctoral funding during the majority of this work, as well as the Chaire of Université Paris-Saclay for providing computational resources. Kavli IPMU is supported by the World Premier International Research Center Initiative (WPI), MEXT, Japan.
\end{acknowledgements}

\pagebreak

\end{document}